\documentclass[a4paper]{aa}
\usepackage{psfig}
\usepackage{epsfig}
\usepackage{graphicx}

\def \inte {INTEGRAL}
\def \xmm {XMM--Newton}

\def \src {SAX~J2103.5+4545}

\def \hcm {\hbox {\ifmmode $ atom cm$^{-2}\else atom cm$^{-2}$\fi}}

\def \pdot {\.P}

\def\approxgt{\mathrel{\hbox{\rlap{\lower.55ex \hbox {$\sim$}}
        \kern-.3em \raise.4ex \hbox{$>$}}}}
\def\approxlt{\mathrel{\hbox{\rlap{\lower.55ex \hbox {$\sim$}}
        \kern-.3em \raise.4ex \hbox{$<$}}}}

\newcommand {\Msun}{M_\odot}
\begin{document}

%\thesaurus{(02.01.2; 08.09.2; 08.14.1; 10.07.3; 13.25.2; 13.25.3)}

\title{A large spin--up rate  measured  with \inte\ in the High Mass X--ray Binary Pulsar \src}

\author{L. Sidoli\inst{1}
        \and S. Mereghetti\inst{1}
        \and S. Larsson \inst{2}    
	\and M. Chernyakova \inst{3}
	\and I. Kreykenbohm \inst{4,3}
	\and P. Kretschmar \inst{5,3}
	\and A.~Paizis \inst{1,3}
	\and A. Santangelo \inst{4}
	\and C. Ferrigno \inst{6}
        \and M. Falanga \inst{7}
}
\offprints{L.Sidoli (sidoli@mi.iasf.cnr.it)}

\institute{
        Istituto di Astrofisica Spaziale e Fisica Cosmica --
        Sezione di Milano -- IASF/INAF, I-20133 Milano, Italy
\and
	Department of Astronomy, Stockholm University, SE-10691 Stockholm, Sweden
\and
	INTEGRAL Science Data Center, 1290 Versoix, Switzerland 
\and   
	Institut f\"ur Astronomie und Astrophysik -- Astronomie,
                 Univ. of T\"ubingen, 72076 T\"ubingen, Germany   
\and
        ESA -- European Space Astronomy Center, E-28080 Villafranca del Castillo, Spain
\and
	Istituto di Astrofisica Spaziale e Fisica Cosmica --
        Sezione di Palermo -- IASF/INAF, I-90146 Palermo, Italy
\and 
        CEA Saclay, DSM/DAPNIA/Service d'Astrophysique (CNRS FRE 2591), F-91191, Gif sur Yvette, France
}

\date{Received 1 March 2005; Accepted: 26 May 2005 }

\authorrunning{L. Sidoli et al.}

\titlerunning{{Large spin-up in  \src\ with \inte}}

\abstract{ 
The High Mass X--ray Binary Pulsar \src\ has been observed with INTEGRAL several times during
the last outburst in 2002--2004.
We report a comprehensive study of all INTEGRAL observations,
allowing a study of the pulse period evolution during the recent outburst. 
We measured a very rapid spin-up episode, lasting $\sim$130~days, which decreased the pulse period
by $\sim$1.8~s. 
The spin-up rate, \pdot$\sim$$-1.5\times$10$^{-7}$~s~s$^{-1}$, 
is the largest ever measured for \src, and it is among the
fastest for an accreting pulsar. The pulse profile shows evidence for
temporal variability, apparently not related to the source flux 
or to the orbital phase. 
The X--ray spectrum is hard and there is significant emission up to
150\,keV.  A new derivation of the orbital period, based on RXTE data, is also reported.
}
\maketitle

\section{Introduction}
\label{sect:intro}

SAX~J2103.5$+$4545 is a transient Be High Mass X-ray Binary 
(HMXRB) pulsar (pulse period $\sim$358\,s) discovered with the 
Wide Field Cameras (WFC) on-board BeppoSAX during  an  
outburst  in February--September 1997  (Hulleman et al., 1998).
A peak intensity of 20\,mCrab was reached on 1997 April 11.
The X--ray spectrum in the 2--25\,keV energy band 
was fit by a power-law with a photon index $\Gamma$=1.27$\pm{0.14}$
modified by photo electric absorption at lower energies
($N_{\rm H}$=3.1$\times$10$^{22}$ cm$^{-2}$).

An orbital period of 12.68\,days and an orbital eccentricity, $e$, of
0.4$\pm$0.2 have been measured with the Rossi X-ray Timing Explorer
(RXTE) during a second outburst lasting about 14
months observed in 1999  (Baykal et al., 2000).

The likely optical counterpart is a B0Ve star (V=14.2; Reig \&
Mavromatakis, 2003; Reig et al. 2004), making \src\ the Be/X-ray
binary with the shortest orbital period known.  \src\ does not follow
the correlation found by Corbet (1986) between the orbital and spin
periods in Be/X-ray binaries (which would imply an orbital period
$\sim$190\,days).

The absorption in the direction of the optical counterpart indicates a
distance of 6.5\,kpc (Reig et al., 2004), doubling the previous
estimate of $\sim$3.2\,kpc (Baykal et al., 2002) based on a model for
the spin-up produced by accretion torques (Ghosh \& Lamb, 1979).
 
A transition from the spin-up phase to the spin-down regime was
observed with RXTE during the decline in the X-ray flux of the 1999
outburst (Baykal et al., 2002).  During the initial part of
the outburst the source exhibited a spin-up phase (the pulse period
decreased by $\sim$0.9\,s in 150\,days), but when the flux dropped the
pulse frequency saturated and, as the flux continued to decline, a
weak spin-down phase started.
This correlation between spin-up rate and X-ray flux indicates the
formation of an accretion disk during the periastron passage.  The
X-ray spectrum, described by a power-law with a cut-off around 8\,keV,
displayed an iron emission line at 6.4\,keV (Baykal et al., 2002).

During a XMM-Newton observation performed in January 2003, Inam et al.
(2004), detected for the first time a soft blackbody component with a
temperature of 1.9\,keV and radius 0.3\,km (assuming a 3.2\,kpc
distance), probably produced in the polar cap of the neutron star, as
well as a transient quasi-periodic oscillation at 22.7\,s.  
Besides the blackbody component, the 1--20 keV spectrum (using \xmm\ 
and RXTE data) was fit with a power-law with a cut-off at 7.89 keV and
an e-folding energy of 27.1 keV, together with an emission line at
6.42\,keV. 
The equivalent width of this cold iron line was found to depend on the
spin phase.  Using RXTE data (MJD\,52611.48 -- MJD\,52668.90), Inam et
al. (2004) derived a pulse period of 354.7940$\pm{0.0008}$\,s (at the
epoch of the XMM-Newton observation, MJD\,52645.85) and a 57.5\,days
average spin-up rate of (7.4$\pm{0.9}$)$\times$10$^{-13}$\,Hz\,s$^{-1}$.

Analyses of part of the INTEGRAL observations of the source have been
reported by Lutovinov et al. (2003), Sidoli et al. (2004), Blay et al.
(2004), Filippova et al. (2004) and Falanga et al. (2005).
In this paper we report on the timing and spectral analysis of all
\inte\ observations performed between December 2002 and July 2004.

\section{Observations}
\label{sect:obs}

The ESA INTEGRAL observatory, launched in October 2002, carries 3
co-aligned coded mask telescopes: the imager IBIS (Ubertini et al.
2003), which allows high-angular resolution imaging in the energy
range 15\,keV--10\,MeV, the spectrometer SPI (Vedrenne et al. 2003; 20
keV--8\,MeV) and the X-ray monitor JEM-X (Lund et al. 2003;
3--35\,keV).  IBIS is composed of a low-energy CdTe detector ISGRI 
(Lebrun et al. 2003), sensitive in the energy range from 15\,keV to
1\,MeV, and a CsI detector (PICsIT; Labanti et al. 2003), designed
for optimal performance at 511 keV, and sensitive in the
175\,keV--10\,MeV energy range.

The INTEGRAL satellite performs regular scans of the Galactic Plane
(GPS) every 12\,days as part of the Core Programme to monitor the
timing and spectral properties of known  X--ray sources,
discover new transient sources and map the diffuse emission
from the Galactic plane (Winkler et al. 2003).

From December 2002 to July 2004, \src\ has been observed several
times.  Among these observations (170 pointings), some have been 
performed during the satellite performance and verification phase
(December 2002), during the GPS (January 2003--July 2004), or during 
guest observer observations now in the public archive (May 2003).  We
have performed a comprehensive study of all these observations (see
Table~\ref{tab:allobs}), concentrating in particular on the timing
properties of \src.

The data reduction has been performed using the OSA~4.2 
release of the \inte\ analysis software, with the corresponding
response matrices.  Each pointing has a typical exposure of
$\sim$2--3\,ks. Only observations where the \src\ position offset was
less than 9$^\circ$ have been considered.  
Only for a few of these
pointings also JEM-X data were available, due to the smaller field of
view of this instrument (offset less than 3.5$^\circ$).  \src\ has
been detected in the energy range 20--40\,keV with IBIS/ISGRI in 113
pointings (see Table~\ref{tab:allobs}).

\begin{figure}[!ht]
\centerline{\psfig{figure=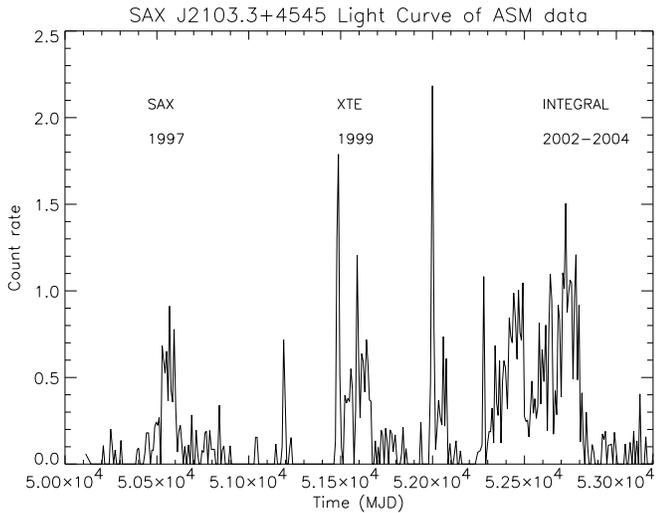,height=6.8cm,angle=0}}
\vskip 0.0truecm
\caption{ASM/RXTE lightcurve (8-days average) 
of SAX~J2103.5+4545 from the outburst in 1997 to the end of August 2004. 
Three distinct outbursts
are evident, and have been marked as ``SAX 1997'' (time of the source discovery
with BeppoSAX, Hulleman et al. 1998), ``XTE 1999'' (observations with XTE 
reported by Baykal et al. 2000), ``2002--2004'' outburst observed with \inte.
}
\label{fig:lc}
\end{figure}

The 2--10\,keV lightcurve of \src\ obtained with the RXTE All Sky
Monitor (ASM) from 1997 to 2004 is shown in Fig.~\ref{fig:lc}.  The
times of the two previous outbursts observed with BeppoSAX (in 1997)
and RXTE (in 1999), as well as the epoch of the INTEGRAL observations
reported here, are indicated.

%%%%%%%%%%%%%%%%%%%%%%%%%%%%%%%%%%%%%%%%%%%%%%%%%%
%%% TAB of all the observations
%%%%%%%%%%%%%%%%%%%%%%%%%%%%%%%%%%%%%%%%%%%%%%%%%%

%%------TABLE 1----------------------------------------------------
\vspace{1.5cm}
\begin{table}[ht!]
\caption{Summary of the \inte\ observations of \src. 
The observations have been grouped together for brevity. 
The 1st and 2nd columns
list the Start and Stop Time  
of each group of observations, 
column n.3 reports the number of ISGRI pointings 
 where the source 
is within 9$^\circ$ of the pointing direction,
column n.4 reports the number of ISGRI/IBIS pointings where \src\ has been detected  in
the energy range 20--40\,keV, 
while the last column
shows if a search for periodicity has been succesfully carried out.
}
%
%\vspace{0.5cm}
\label{tab:allobs}
\begin{center}
\begin{tabular}[c]{llccc}
\hline\noalign{\smallskip}
Start Time             &  End Time        &   Number   &  Detections                & Timing   \\ 
 (MJD)                 & (MJD)            &   of obs.  &                           &     \\
\noalign{\smallskip\hrule\smallskip}
52617.86   &	52619.28	&	4	& 	3	        &	N   \\  
52620.74   &    52620.93	&	2	&	0		&	N  \\
52629.63   &    52631.92	&	58	&	56              &       Y  \\
52636.66   &    52637.87	&    	21	&	0		&	N \\
52638.46   &    52640.07	&	2	&	0		&	N \\
52653.32   &	52653.40	&	2	&	1               &       N   \\	
52722.85   &	52722.93	&	3	&	0		&	Y   \\
52737.03   &	52737.12	&	3	&	1               &       Y   \\
52746.10   &    52746.18	&	3	&	3               &       Y   \\
52761.26   &    52762.42  	& 	38	&	38		&       Y   \\
52770.05   &    52770.10        &       2       &       2     	        &	Y   \\
52782.06   &    52782.14        &       3	&	3               &       Y   \\
52835.9    &	52836.03	&	3	&       1               &       N  \\
52985.41   &    52985.49        &       3       &       0               &       N  \\
52994.38   &    52994.47        &       3       &       0               &       N  \\
53019.37   &    53019.45        &       3        &      0               &       N  \\
53045.61   &    53045.69        &       3        &      0               &       N  \\
53090.96   &    53091.00        &	1	&	0               &      N  \\
53102.48   &	53102.56	&	3	&	0               &      N   \\
53114.16   &	53114.24	&	3	&	0               &      N   \\
53126.09   &    53126.14        &       2      &	2		&	Y \\
53137.88   &	53137.93	&	2	&	0               &     N   \\
53164.97   &	53165.05	&	3	&	0                &     N   \\
53188.96  &   	53189.03	&	2	&	0               &     N   \\
\noalign{\smallskip\hrule\smallskip}
\end{tabular}
\end{center}
\end{table}

%%%%%%%%%%%%%%%%%%%%%%%%%%%%%%%%%%%%%%%%%%%%%%%%%%
%%%%%%%%%%%%%%%%%%%%%%%%%%%%%%%%%%%%%%%%%%%%%%%%%%%%%%%%%%%%%%%%%%%%%%%%%%%%%%%

\section{Results}

\subsection{Timing Analysis}

In order to perform the timing analysis, we selected events with a 
Pixel Illumination Function (PIF, pixel fraction illuminated
by the source) larger than 0.5. 
IBIS/ISGRI events have been collected
in the energy band 20--40\,keV, where the best statistics is available, 
while for JEM-X the  range 3--30\,keV has been used.

\begin{figure}[!ht]
\centerline{\psfig{figure=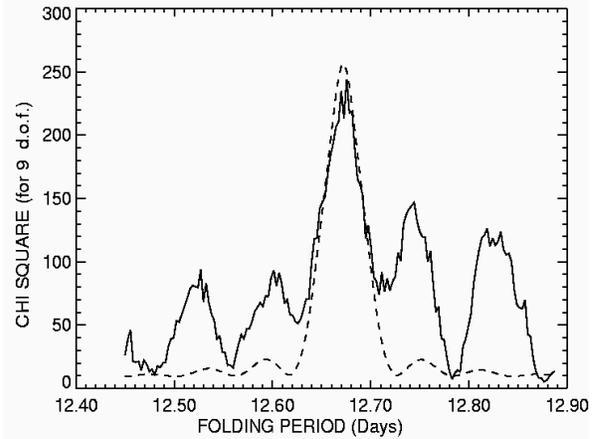,height=6.0cm,angle=0}}
\vskip 0.0truecm
\caption{$\chi^2$ distribution using all avilable RXTE-ASM data. 
It peaks at 12.670$\pm{0.005}$\,days, our improved determination of the orbital
period for \src.  The solid line marks the data and the 
dashed line the fitted model 
(see text).
}
\label{fig:orb_chi}
\end{figure}

\begin{figure}[!ht]
\centerline{\psfig{figure=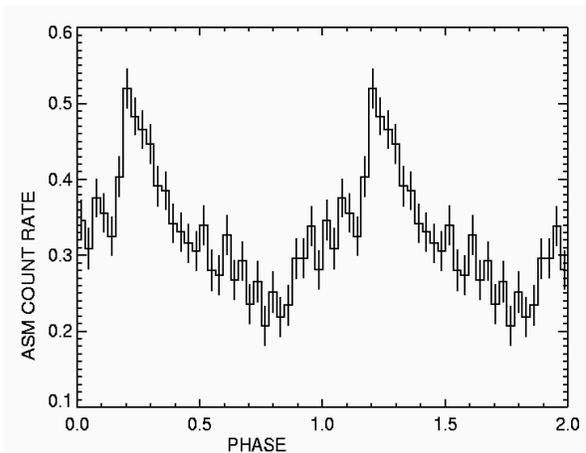,height=6.0cm,angle=0}}
\vskip 0.0truecm
\caption{SAX~J2103.5+4545 RXTE-ASM lightcurve folded at the updated
orbital period of 12.670$\pm{0.005}$\,days.
}
\label{fig:orb_folded}
\end{figure}

Arrival times have been corrected to the Solar System barycenter and
for the Doppler delay due to the motion in the binary system.  To
perform this orbital correction, we adopted the ephemeris zero point from Baykal
et al. (2000), but used an updated orbital period, 
P$_{\rm orb}$=12.670$\pm{0.005}$\,days, that we
calculated with epoch folding of all
the available RXTE/ASM data (see Fig.~\ref{fig:orb_chi} and
Fig.~\ref{fig:orb_folded}). 
For our analysis, based on ASM dwell-by-dwell
data, we used the epoch folding method described in Larsson (1996). 
The error estimate is based on Monte Carlo simulations and on 
the analysis of pulsations injected into the ASM lightcurve.
From the Monte Carlo simulations we estimate the statistical error 
to be  $\pm{0.003}$~days. The slightly higher error estimate
above includes also effect of the irregular source variability
on the period determination. This latter effect is estimated
from the injected pulsation.

%%%%%%%%%%%%%%%%%%%%%%%%%%%%%%%%%%%%%%%%%%%%%%%%%%%%%%%%%%%%%%%%%%%%%%%%%%%%
%%%%%%%%%%%%%%%%%%%%%%%%%%%%%%%%%%%%%%%%%%%%%%%%%%%%%%%%%%%%%%%%%%%%%%%%%%%%
%%%%%%%%%%%%%%%%%%%%%%%%%%%%%%%%%%%%%%%%%%%%%%%%%%%%%%%%%%%%%%%%%%%%%%%%%%%%

Since the short duration of each single \inte\ pointing
($\sim$2--3\,ks) leads to a large uncertainty in the
determination of the pulse period (typically about 10\,s, even in
presence of good statistics), we grouped the observations where a good
estimate of the pulse period could be achieved in 7 separate
data-sets consisting of the pointings performed during the same
satellite revolution. 

In Table~\ref{tab:obslog} we list the start and end time of each group
of observations, together with the source average IBIS/ISGRI count
rate in the 20--40\,keV range.  In data-sets n.1, n.4 and n.5 also
JEM-X data are available, and have been used in the timing analysis.

%%------TABLE 1----------------------------------------------------
\vspace{1.5cm}
\begin{table*}[ht!]
\caption{Results of the \src\ timing analysis.
The 2nd and 3rd columns
list the Start and Stop Time  
of each data-set, the 4th the averaged IBIS/ISGRI (20--40 keV) 
source count rate  in these epochs, the 5th and 6th columns list 
the pulse periods measured 
in the seven \inte\ data-sets. In data-sets n.2, 3, 6 and 7
the source was outside the JEM-X field of view. In data-set n.1 only
the pulse period estimated with JEM-X is reported due to 
several telemetry gaps present in ISGRI lightcurve during this satellite
revolution, which hampered a proper
timing analysis
}
\label{tab:obslog}
\begin{center}
\begin{tabular}[c]{cllcr@{.}lc}
\hline\noalign{\smallskip}
Data-set  & Start Time             &  End Time        &   IBIS/ISGRI        &  \multicolumn{3}{c}{Pulse Period (s)}      \\  
 ID.              & (MJD)                 & (MJD)            &  (s$^{-1}$)  &  \multicolumn{2}{c}{  IBIS/ISGRI}       &   JEM-X         \\
\noalign{\smallskip\hrule\smallskip}
1 &           52629.91           &        52631.84  &    5.16$\pm{0.20}$ &     \multicolumn{2}{c}{  -- }         &  355.09 $\pm{0.03}$    \\
2 &           52722.85           &        52722.90  &    4.28$\pm{0.18}$  &    356&\phantom{33} $\pm{3}$   & $-$    \\
3 &           52746.10           &        52746.18  &    6.07$\pm{0.25}$   &  355&5\phantom{3} $\pm{1.0}$ & $-$   \\
4 &           52761.26           &        52762.42  &    4.07$\pm{0.08}$   &   353&32 $\pm{0.03}$ & 353.47 $\pm{0.15}$    \\
5 &           52770.05           &        52770.10  &    8.62$\pm{0.27}$   &   351&8\phantom{3} $\pm{1.3}$      & 355.4 $\pm{2.3}$   \\
6 &           52782.06           &        52782.12  &    6.45$\pm{0.28}$  &    351&3\phantom{3} $\pm{2.3}$      & $-$   \\
7 &           53126.09           &        53126.14  &    4.34$\pm{0.28}$  &    355&\phantom{33} $\pm{3}$          & $-$   \\
\noalign{\smallskip\hrule\smallskip}
\end{tabular}
\end{center}
\end{table*}

%%%%%%%%%%%%%%%%%%%%%%%%%%%%%%%%%%

\vskip -1.5truecm
\begin{figure}[!ht]
\centerline{\psfig{figure=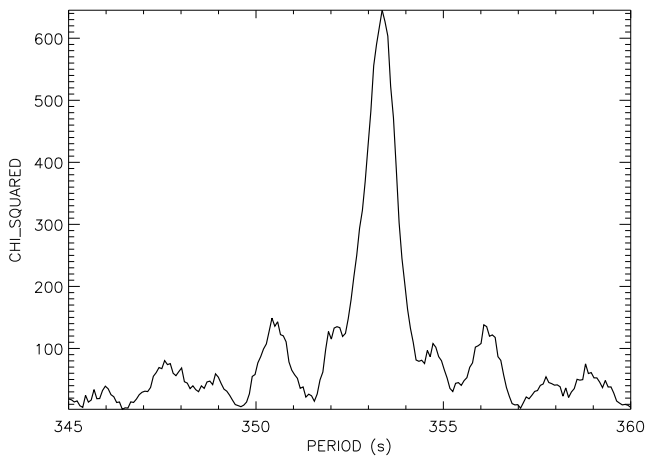,height=6.5cm,angle=0}}
\vskip 0.0truecm
\caption{Distribution of $\chi^2$ versus trial period for 
data-set n.4 (see Table~\ref{tab:obslog}) 
in the energy range 20--40\,keV
}
\label{fig:chi}
\end{figure}

%%%%%%%%%%%%%%%%%%%%%%%%%%%%%%%%%%

We searched for the \src\ spin period by using epoch 
folding techniques:
we produced $\chi^2$ distributions versus trial period,
and then we fit them in order to derive the best-fit period for each
data-set, and its associated uncertainty (Leahy 1987).  An example of
a $\chi^2$ distribution is shown in Fig.~\ref{fig:chi}.

The derived pulse periods are reported in Table~\ref{tab:obslog}.
Our data-sets n.1 and n.4 include observations partly reported by  
Blay et al. (2004), Filippova et al. (2004) 
and Falanga et al. (2004).

A period difference of $\sim$1.77\,s
in $\sim$130.5\,days is evident from the comparison of results from data-sets n.1
and n.4, which have the smallest uncertainty
(Table~\ref{tab:obslog}). 
This spin-up is significantly higher than that measured during 
previous  outbursts of this source.

\begin{figure*}[!ht]
\centerline{\psfig{figure=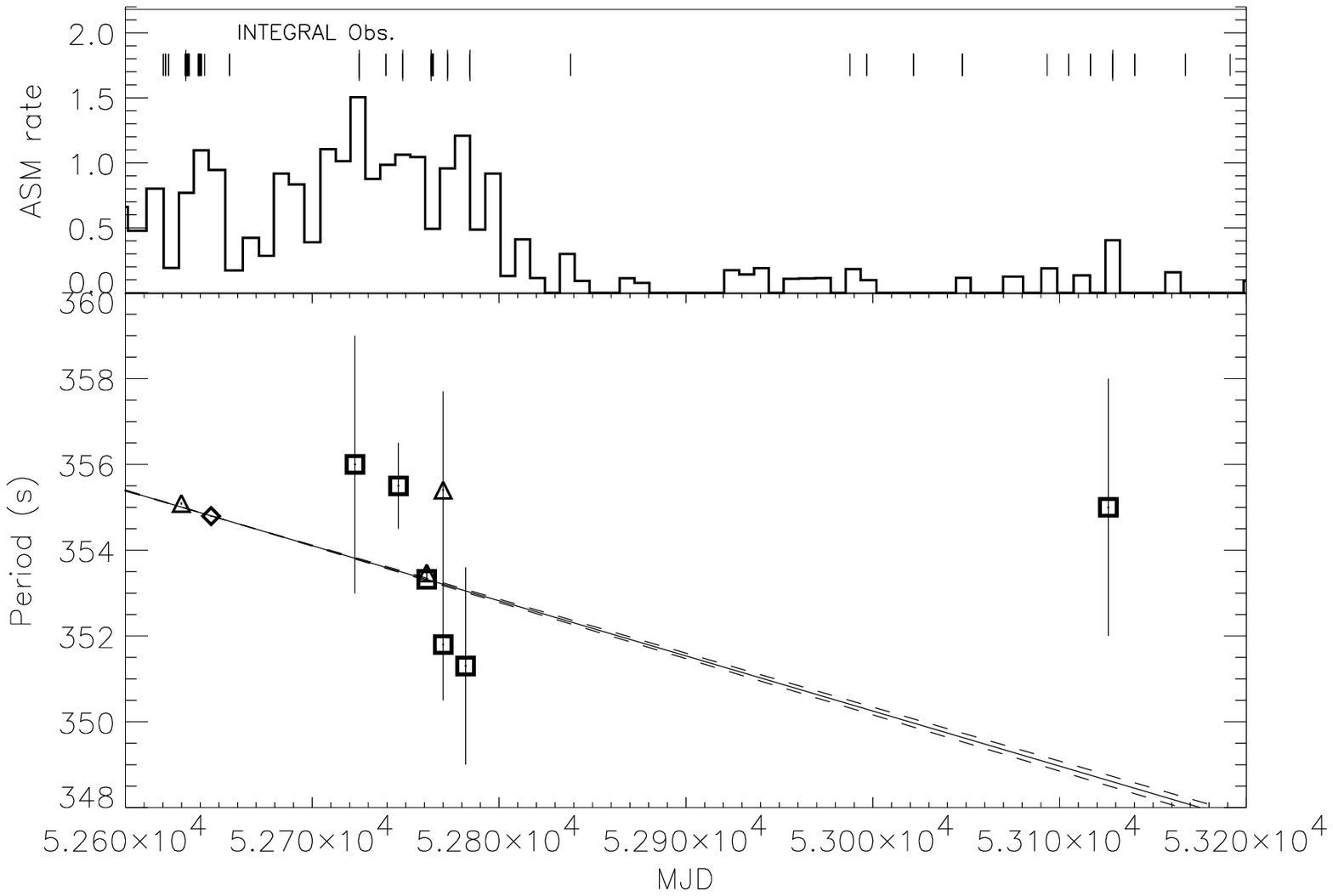,height=12cm,angle=0}}
\vskip 0.0truecm
\caption{{\em Upper panel}: A close-up view of the ASM/RXTE lightcurve (8-days average)
  during the INTEGRAL observations. The epochs of the INTEGRAL 
  observations have been marked with vertical lines \emph{Lower
    panel}: SAX~J2103.5+4545 pulse period evolution (triangles mark
  JEM-X results; squares IBIS/ISGRI results).  The diamonds indicate
  the period measured with XTE by Inam et al. (2004).  A linear fit to
  all the data is reported (solid line) together with the
  1-$\sigma$~uncertainty (dashed lines).  }
\label{fig:spinup}
\end{figure*}

The pulse period evolution during the \inte\ observations is shown in
Fig.~\ref{fig:spinup}, where also the period measured by Inam et al.
(2004) with RXTE has been shown for completeness.  A linear fit to all
these data indicates an average spin-up with a rate of
\pdot=$-$$(1.487\pm{0.030})\times$10$^{-7}$\,s\,s$^{-1}$ ($\chi^2$=21.3,
8 degrees of freedom, dof).

\begin{figure*}
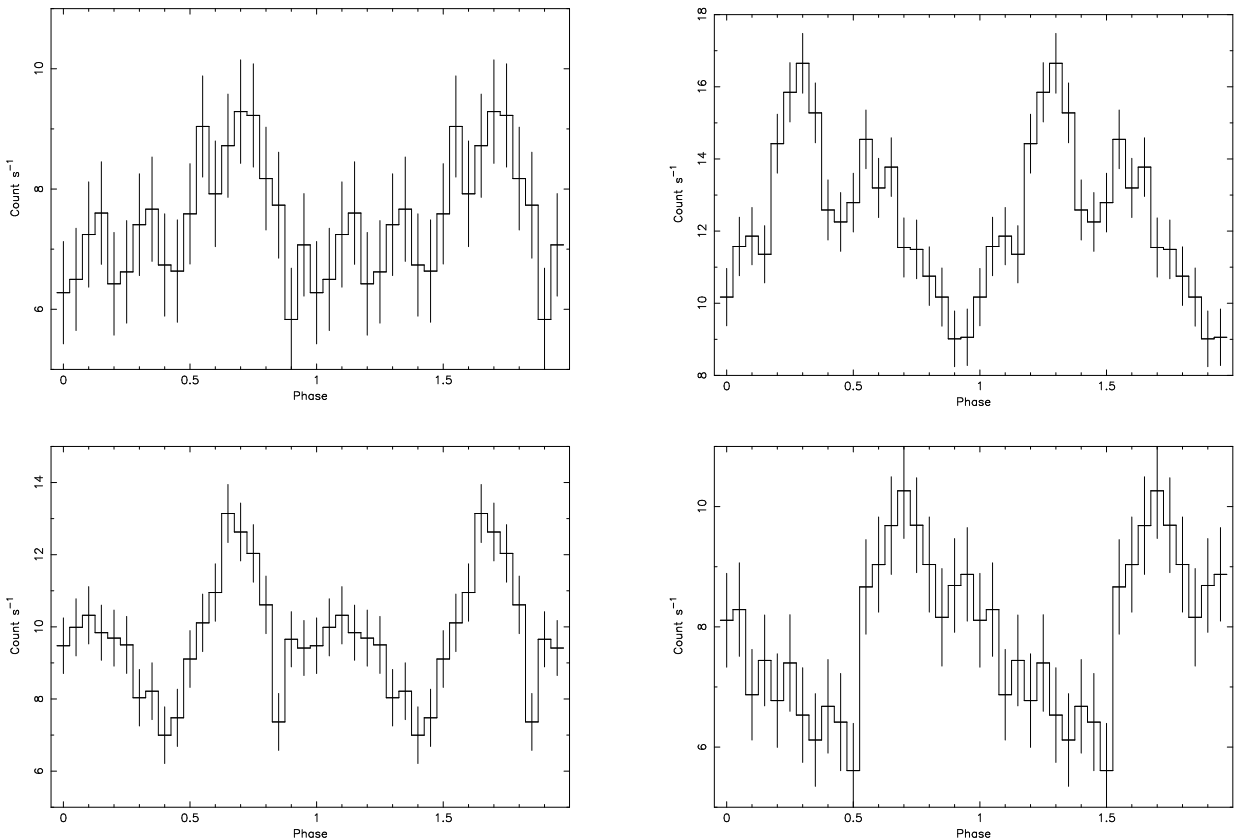

\hbox{\hspace{0.6cm}
\includegraphics[height=7.5cm,angle=-90]{aa20052961_fig3a.ps}
\hspace{1.1cm}
\includegraphics[height=7.5cm,angle=-90]{aa20052961_fig3b.ps}}
\vbox{\vspace{0.1cm}}
\hbox{\hspace{0.6cm}
\includegraphics[height=7.5cm,angle=-90]{aa20052961_fig3c.ps}
\hspace{1.1cm}
\includegraphics[height=7.5cm,angle=-90]{aa20052961_fig3d.ps}}
\caption[]{Few examples of \src\ pulse profiles
from \inte\ observations where a different
shape is evident.
From top to bottom, from the left to the right, profiles are 
from data-sets n.2, n.5, n.6 and n.7, all  in the energy range 15--40\,keV.
The zero phase is arbitrary.
The pulsed fraction, defined as the ratio of maximum minus minimum 
to maximum, is $\sim$33\% for data-set n.2, while $\sim$47\%--50\% for the other three
lightcurves
}
\label{fig:profiles}
\end{figure*}

\begin{figure*}
\hbox{\hspace{0.01cm}
\includegraphics[height=6.0cm]{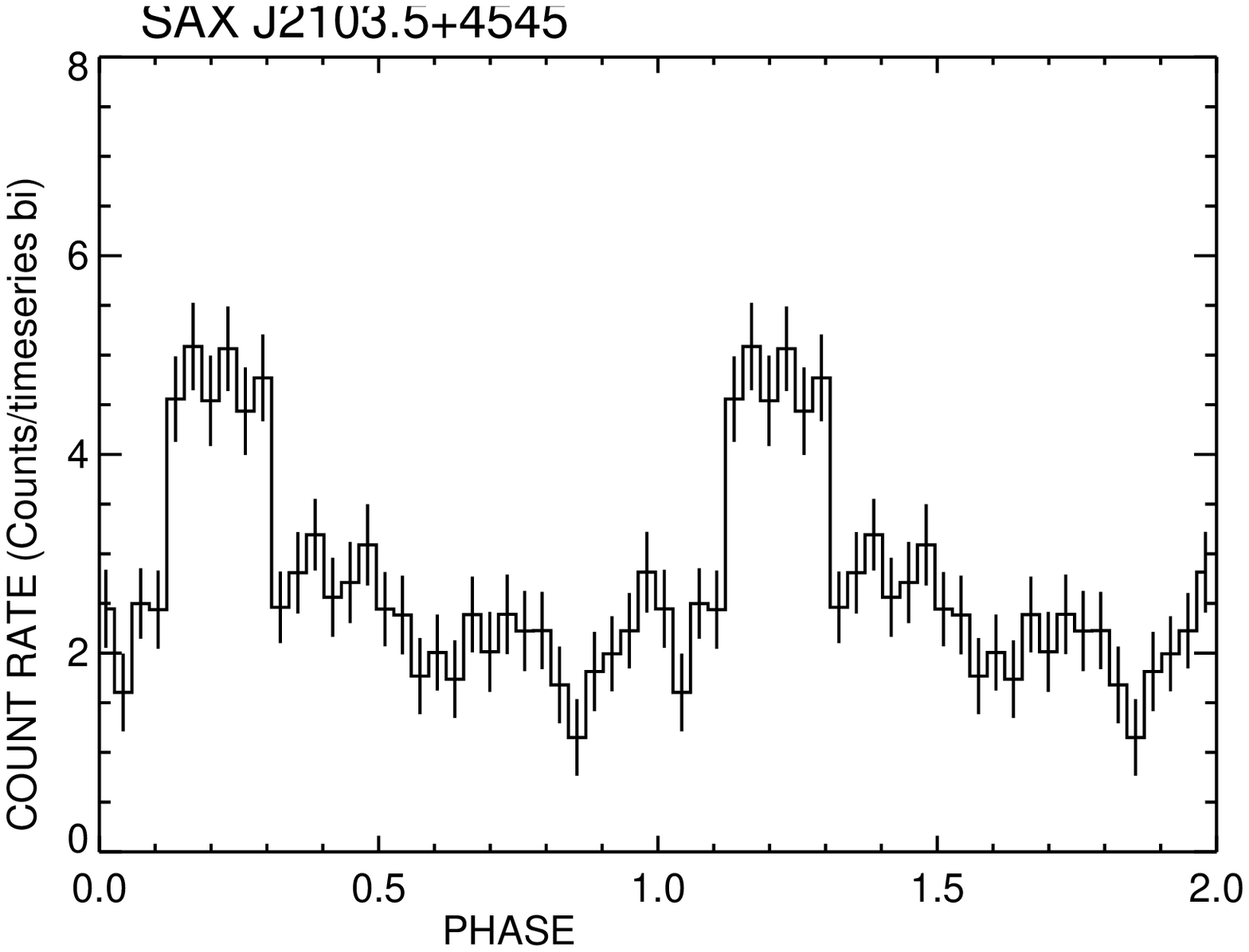}
\hspace{0.2cm}
\includegraphics[height=6.0cm]{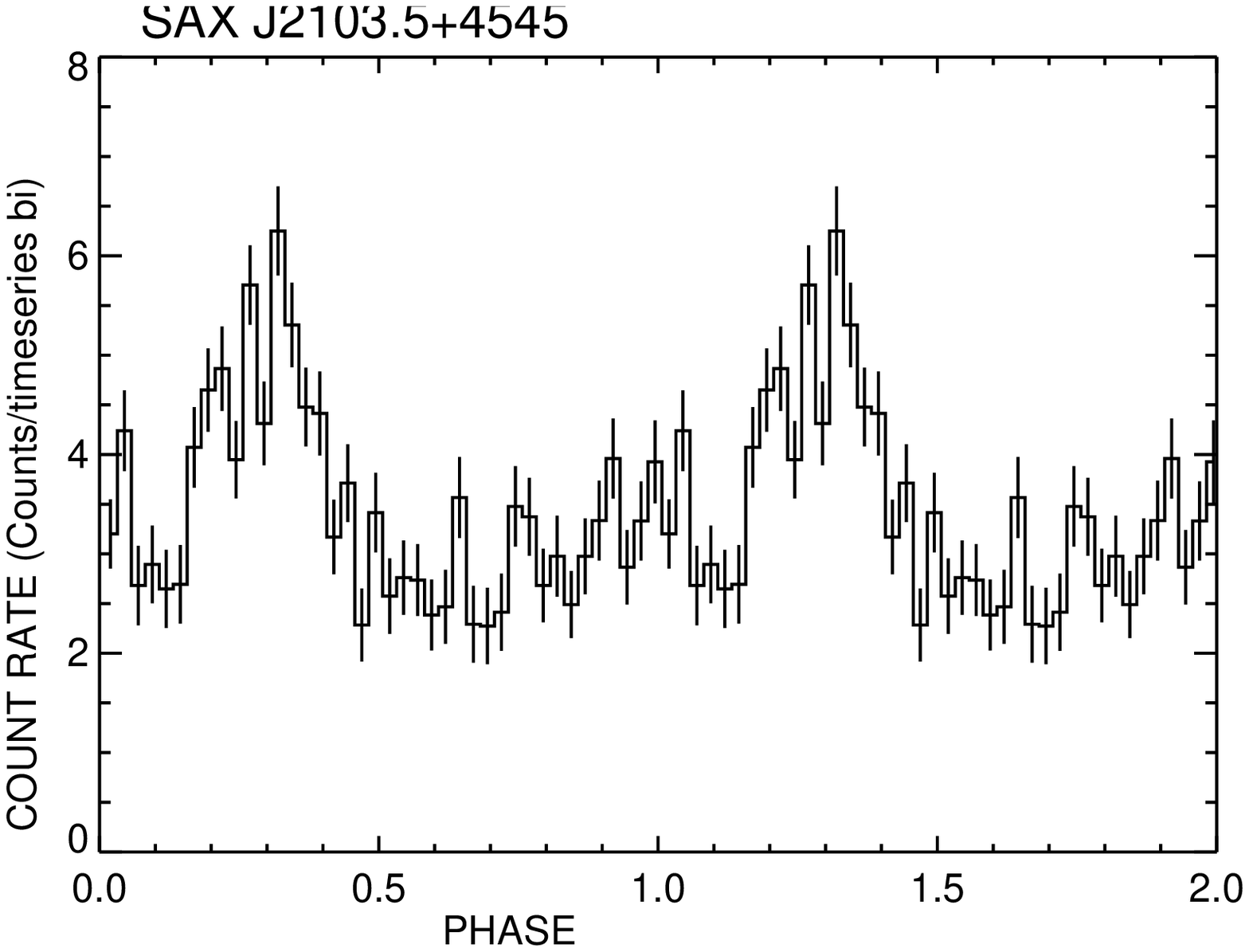}}
\caption[]{Pulse profiles 
at soft energies (3--30\,keV) with JEM--X,
from data-set n.1 (left panel) and n.4 (right panel). 
}
\label{fig:jemx_profiles}
\end{figure*}

A few examples of the folded light curves (background subtracted)
obtained in different observations with ISGRI and JEM-X are shown in
Fig.~\ref{fig:profiles} and Fig.~\ref{fig:jemx_profiles},
respectively.

%%%%%%%%%%%%%%%%%%%%%%%%%%%%%%%%%%%%%%%%%%%%%%%%%%%%%%%%%%%%%%%%%%%%%%%%%%% 


%%%%%%%%%%%%%%%%%%%%%%%%%%%%%%%%%%%%%%%%%%%%%%%%%%%%%%%%%%%%5

%%%%%%%%%%%%%%%%%%%%%%%%%%%%%%%%%%%%%%%%%%%%%%%%%%%%%%%%%%%%%%%%%%%%%%%%%%%%

\subsection{Spectral Analysis}

We used data from all \inte\ observations to derive a spectrum with a
very high signal to noise ratio. Since the source exhibits no
significant spectral evolution during the outburst, this can be safely
done.

The average source spectrum from 5 to 200\,keV, has been investigated
by simultaneously fitting JEM--X (5--20\,keV) together with IBIS/ISGRI
(20--200\,keV), and SPI (20--200\,keV) spectra, achieving net exposure
times of 40\,ks, 290\,ks and 130\,ks, respectively for the three
instruments.  An additional uncertainty of 5\% was added quadratically
to the statistical errors to account for systematic uncertainties in
instrumental responses.  Free relative normalizations between the
three instruments were included.  Since the energy range (5--200 keV)
is not sensitive to the interstellar column density, we
fixed it at 3$\times$10$^{22}$\,cm$^{-2}$ (Hulleman et al., 1998).

\begin{figure*}
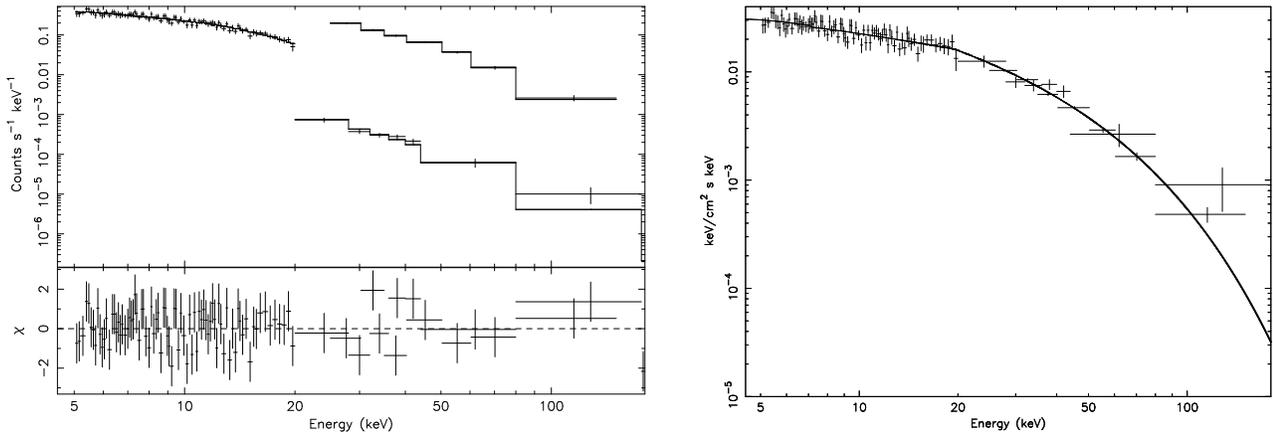

\hbox{\hspace{.3cm}
\includegraphics[height=8.5cm,angle=-90]{aa20052961_fig6a.ps}
\hspace{.5cm}
\includegraphics[height=7.6cm,angle=-90]{aa20052961_fig6b.ps}}
\caption[]{{\em Left panel:}  SAX~J2103.5+4545 counts spectrum (5--200 keV) 
extracted from JEM-X (5--20 keV), IBIS/ISGRI (20--200 keV, upper counts spectrum)
and SPI (20--200 keV, lower spectrum) instruments.
The best-fit model (histogram) is an absorbed
power-law with a cut-off at high energies (see text for the best-fit parameters).
The residuals displayed in the lower panel are in units of standard
deviations. {\em Right panel:} SAX~J2103.5+4545 unfolded spectrum together
with the best-fit model (solid line).
}
 \label{fig:spec}
\end{figure*}

A single power-law is not able to  describe the broad--band 
spectrum (reduced $\chi^2$$>$2).  The best-fit ($\chi^2$=121.9 for 129
d.o.f.)  is an absorbed power-law with a high energy cut-off (see
Fig.~\ref{fig:spec}), with the following parameters: photon index
$\Gamma$=1.53 $^{+0.08} _{-0.07}$, cut-off energy E$_{\rm c}$=19
$^{+11} _{-4}$\,keV and e-folding energy E$_{\rm f}$=32$\pm{3}$\,keV.
The 5--200 keV observed flux is
$\sim$1.05$\times$10$^{-9}$\,erg\,cm$^{-2}$\,s$^{-1}$ (based on the JEM--X
response matrix).
The source flux in the  2--10\,keV energy range, 
corrected for the absorption, is
$\sim$4.2$\times$10$^{-10}$\,erg\,cm$^{-2}$\,s$^{-1}$, which
translates into a luminosity 
L$_{\rm X}$=2$\times$10$^{36}$\,erg\,s$^{-1}$~(2--10\,keV, 
for a distance of
6.5\,kpc).

%%%% new figure after referee's comments:

%%%%%%%%%%%%%%%%%%%%%%%%%%%%%%%%%%%%%%%%%%%%%%%%%%%%%%%%%%%%%%%%%%%%%%%%%%%%%%%%%%%%%%%
\section{Discussion}
\label{sect:discussion}
 
We have studied all the available \inte\ observations of \src, obtained
from December 2002 to July 2004.  
They cover part of the latest source
outburst, which is the longest and brightest observed from this source
since its discovery in 1997.

%%% ------------ > ORBITAL PERIOD
To perform a proper timing analysis of the \inte\ observations, 
we re-determined the orbital period of the binary system using ASM/XTE public data, 
obtaining a refined P$_{\rm orb}$ of 12.670$\pm{0.005}$\,days
(to be compared with P$_{\rm orb}$ of 12.68$\pm{0.25}$\,days, measured
with XTE during the 1999 outburst by Baykal et al., 2000).

%%% ------------ > PULSE PROFILES
The \src\ pulse profiles observed with \inte\ show evidence for
temporal variability over the same energy range (15--40 keV),
where a single asymmetric pulse profile seems to evolve to a double-peaked profile
in some of the observations. 
On the other hand, in the lightcurves extracted
at softer energies (3--30~keV) there is no clear evidence for a secondary peak.
The \inte\ observations unfortunately do not cover the
entire orbital phase; in particular, the data-sets n.3, n.5, n.6 and
n.7 sample roughly the same orbital phase (these are indeed GPS
observations, which are performed every 12\,days, a value close to the
source orbital period).  Thus, apparently the complex variability of the pulse
profile does not depend on the orbital phase. Also an obvious dependence on the
source luminosity seems to be unlikely, since we observe different pulse
periods from observations at similar fluxes (e.g., data-sets n.2 and n.7).
The complex variability of the pulse profiles is a common feature among
the class of accreting binary pulsars, also over the same energy range
(White, Swank \& Holt, 1983; Bildsten et al., 1997). In some sources the
pulse shape variability shows a clear dependence on the source luminosity
(first observed in EXO~2030+375, Parmar et al., 1989) or on the
energy range (e.g. Vela X--1, Kretschmar et al, 1997).
A simple geometric model with two pencil beams coming from the magnetic poles
is too simple to explain the pulse profiles in most pulsars, and a more
complex emission pattern or the inclusion of scattering effects in
the transfer through highly magnetized plasmas are needed (e.g., Nagel, 1981).
A computation of pulse profiles of medium-luminosity (L$_{\rm X}\sim$10$^{36}$~erg~s$^{-1}$) 
binary X--ray pulsars has been performed by Kraus et al. (2003), who found that
energy-dependent peaks in the profiles are mainly due to energy-dependent
relative importance of the halo (which forms around the accretion funnel where
the neutron star surface is irradiated) and the column contributions to the observed flux.
It is also possible that the temporal variability of 
the pulse profile we observe in the same energy
range can be due to time-dependent emission pattern, 
or to changes in the opacity
of the magnetized plasma where the radiation propagates.

%%% ------------ > SPIN UP
The pulse period evolution measured with \inte\ shows clear evidence for a spin-up
episode, lasting at least $\sim$130\,days from MJD~52629.9 (data-set n.1) 
to MJD~52761.3 (data-set n.4), where the pulse period decreased by $\sim$1.77\,s.

Considering  the overall pulse period history 
during the latest outburst (Fig.~\ref{fig:spinup}), 
a linear fit 
results in a spin-up rate \pdot=$-(1.487\pm{0.030})$$\times$10$^{-7}$\,s\,s$^{-1}$.

We cannot exclude a more complex evolution of the pulse period than a
simple steady spin-up  over more than 1.3\,years. In particular,
it is likely that the spin-up trend induced by the accretion does not
extend until the epoch of the data-set n.7. 

A variation in the accretion rate can modify the pulse period
derivative.  A correlation of the spin-up rate with the X--ray flux has
been indeed found for \src\ by Baykal et al. (2002) during the 1999
outburst, as predicted by the Ghosh and Lamb (1979) model for the
torques exerted on the neutron star during the accretion through a
disk.  
The existence of this correlation is indeed indicative of the formation
of an accretion disk around \src\ during that outburst.
The last pulse period measured with \inte\ (data-set n.7) has
been determined after a long faint state lasting about 300\,days
(Fig.~\ref{fig:spinup}, upper panel).  The reduced accretion rate
could have increased the \src\ pulse period (a spin-down phase
followed the spin-up trend during the 1999 \src\ outburst, Baykal et
al. 2002), although the large uncertainty in the pulse period at that
epoch (P=355$\pm{3}$\,s) does not permit to draw firm conclusions.

The formation of an accretion disk in \src\  
during the outburst in 1999 has been established from RXTE data (Baykal et al. 2002). 
Furthermore, the fact that
our estimate of the spin-up rate (the highest ever measured in this source)
has been measured during the brightest outburst observed from \src, indicates
that the  correlation between luminosity and spin-up is still valid.
It is thus tempting to discuss our results on the source spin-up rate 
in the framework of the Ghosh and Lamb (1979) theory.

\begin{figure}[!ht]
\centerline{\psfig{figure=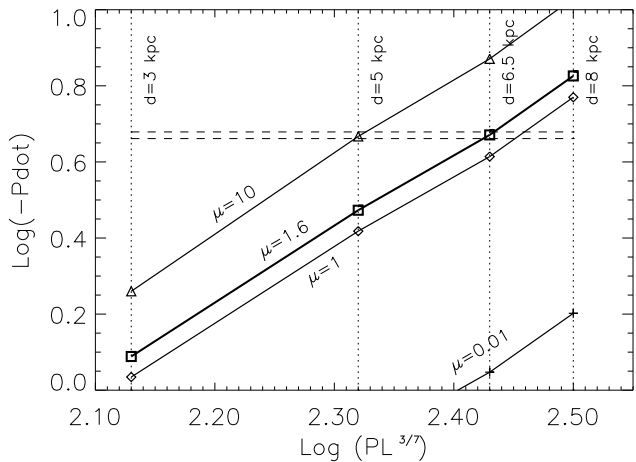,height=6.9cm,angle=0}}
\vskip 0.0truecm
\caption{Measured spin-up rate (horizontal dashed lines show the 1\,$\sigma$ uncertainty 
interval for \pdot=$-(1.487\pm{0.030})$$\times$10$^{-7}$\,s\,s$^{-1}$) 
compared with the theoretical relation
 (from Ghosh \& Lamb, 1979) between the spin-up rate -\pdot (here in units of yr\,s$^{-1}$)
and the quantity  PL$^{3/7}$, 
where P is the pulse period in seconds, and L is the source luminosity in units of
10$^{37}$\,erg\,s$^{-1}$. Each curve has been calculated for a 
different value of the magnetic moment $\mu$
(in units of 10$^{30}$\,Gauss\,cm$^{3}$; 
triangles are 
for 10$\times$10$^{30}$\,Gauss\,cm$^{3}$,
squares for 1.6$\times$10$^{30}$\,Gauss\,cm$^{3}$,
diamonds for 10$^{30}$\,Gauss\,cm$^{3}$, and 
crosses for 
0.01$\times$10$^{30}$\,Gauss\,cm$^{3}$).
A 1.4 $\Msun$ neutron star, a pulse period P=353.32\,s and 
a source flux of 10$^{-9}$\,erg\,cm$^{-2}$\,s$^{-1}$ 
have been considered.
Vertical dotted lines show loci of
equal source distance d: from left to right, d=3 kpc, d=5 kpc, d=6.5 kpc, d=8 kpc.}
\label{fig:theory}
\end{figure}

In Fig.~\ref{fig:theory} we compare the measured pulse period
derivative of $\sim$$-$1.49$\times$10$^{-7}$\,s\,s$^{-1}$ with the
theoretical relation between the spin-up rate of a 1.4\,$\Msun$
neutron star and the quantity PL$^{3/7}$, where P is the pulse period
in seconds, and L is the source luminosity in units of
10$^{37}$\,erg\,s$^{-1}$ (Ghosh \& Lamb, 1979).  Thus, assuming that
the distance of 6.5\,kpc is correct, a magnetic moment of
$\sim$1.6$\times$10$^{30}$\,Gauss\,cm$^{3}$ is needed to explain the
spin-up rate.  This translates into a magnetic field
of $\sim$10$^{12}$\,Gauss, which would imply a cyclotron line at
energies around $\sim$10-20\,keV, not observed in the source spectrum.
The absence of lines could suggest a much higher neutron star magnetic
field, $\sim$10$^{13}$\,Gauss, but this would imply a smaller
distance, around 5\,kpc.
A distance of 5\,kpc also agrees with the maximum spin-up trend
measured with XTE during the 1999 outburst (Baykal et al., 2002): a
pulse period derivative of $\sim$8.9$\times$10$^{-8}$\,s\,s$^{-1}$
(and pulse period P=358.166\,s) for a source flux of
$\sim$6$\times$10$^{-10}$\,erg\,cm$^{-2}$\,s$^{-1}$ can be explained
with a neutron star of 1.4 $\Msun$ and a magnetic moment
$\sim$10$^{31}$\,Gauss\,cm$^{3}$, located at a distance of about
5\,kpc.

However, note that there are uncertainties in the dimensionless
angular momentum adopted in the accretion torque theory (Li \& Wang,
1996), which could significantly modify the estimate of the source
distance  derived with this method 
(see, e.g., Tsygankov \& Lutovinov (2005), on the binary
pulsar KS\,1947+300).  
Another uncertainty is due to the source flux
we adopted here, which is not the bolometric flux and has been
measured from the average \inte\ spectrum in the energy range
5--200\,keV.
%
%%%%%%%%%%%%%%%%%%%%%%%%%%%%%%%%%%%%%%%%%%%%%%%%%%%%%%%%%%%%%%%%%%%%%%%%%%%%%%%%%%%%%%
%%%%%%%%%%%%%%%%%%%%%%%%%%%%%%%%%%%%%%%%%%%%%%%%%%%%%%%%%%%%%%%%%%%%%%%%%%%%%%%%%%%%%%

%%% SUMMARY ON THE SPIN UP
The spin-up rate measured with \inte\ from \src\ is the largest experienced by this source
(likely because of the higher brightness of the latest outburst) 
and is among the largest found for an X-ray binary pulsar.
The frequent monitoring (as that performed with \inte\ during the GPS) 
of accreting pulsars is essential in the determination
of the pulse period evolution, and crucial to test accretion theories.

%\begin{acknowledgements}

%\end{acknowledgements}

\end{document}